\documentclass[prl,twocolumn,showpacs,preprintnumbers,floatfix,superscriptaddress]{revtex4}

\usepackage{graphicx}

\newcommand{\gev}{\,{\rm GeV}}
\newcommand{\tev}{\,{\rm TeV}}

\begin{document}

\preprint{\hbox{JLAB-THY-07-636, ADP-07-04-T644}}
\title{Testing the Standard Model by precision measurement of the weak charges of quarks}

\author{R.~D.~Young}
\affiliation{Jefferson Lab, 12000 Jefferson Ave., Newport News, Virginia 23606, USA}
\affiliation{Special Research Centre for the Subatomic Structure of Matter, and
             Department of Physics, University of Adelaide, Adelaide SA 5005, Australia}
\author{R.~D.~Carlini}
\affiliation{Jefferson Lab, 12000 Jefferson Ave., Newport News, Virginia 23606, USA}
\affiliation{College of William and Mary, Williamsburg,
Virginia 23187, USA}
\author{A.~W.~Thomas}
\affiliation{Jefferson Lab, 12000 Jefferson Ave., Newport News, Virginia 23606, USA}
\affiliation{College of William and Mary, Williamsburg,
Virginia 23187, USA}
\author{J.~Roche}
\affiliation{Jefferson Lab, 12000 Jefferson Ave., Newport News, Virginia 23606, USA}
\affiliation{Ohio University, Athens, Ohio 45071, USA}

\date{\today}

\begin{abstract}
In a global analysis of the latest parity-violating electron
scattering measurements on nuclear targets, we demonstrate a
significant improvement in the experimental knowledge of the weak
neutral-current lepton--quark interactions at low energy. The
precision of this new result, combined with earlier atomic
parity-violation measurements, places tight constraints on the size of
possible contributions from physics beyond the Standard Model.
Consequently, this result improves the lower-bound on the scale of relevant
new physics to $\sim 1\tev$.
\end{abstract}


\maketitle

%%%%%%%%%%%%%%%%%%%%%%%%%%%%%%%%%%%%%%%%%%%%%%%%%%%%%%%%%%%%%

The Standard Model has been enormously successful at predicting the
outcomes of experiments in nuclear and particle physics. The search
for new physical phenomena and a fundamental description of nature
which goes beyond the Standard Model is driven by two complementary
experimental strategies. The first is to build increasingly energetic
colliders, such as the Large Hadron Collider (LHC) at CERN, which aim
to excite matter into a new form. The second, more subtle approach is
to perform precision measurements at moderate
energies \cite{Bennett:1999pd,Zeller:2001hh,Anthony:2005pm}, where an
observed discrepancy with the Standard Model will reveal the signature
of these new forms of matter \cite{Ramsey-Musolf:2006vr}. Here we show
that the latest measurements of the parity violating electroweak
force \cite{Ito:2003mr,Spayde:2003nr,Maas:2004ta,Maas:2004dh,Aniol:2005zf,Aniol:2005zg,Armstrong:2005hs,Acha:2006my}
constrain the possibility of relevant physics beyond the Standard
Model to the TeV energy scale and beyond.
While the current data sets a much improved bound on the scale of new physics, 
the nature of such 
low-energy tests is that future results will play a 
complementary role in determining the structure of potential new
interactions in the LHC era.

After three decades of experimental tests, the only indication of a
flaw in the Standard Model lies in the recent discovery of neutrino
oscillations~\cite{Ahn:2002up}.  That discovery has renewed interest
in identifying other places where physics beyond the Standard Model
might be found. In this work we report the results of a search for
indirect signatures of new physics through precise measurements at low
energy. This is possible because, within the electroweak theory, one
can rigorously derive a low-energy effective interaction between the
electron and the quarks. Any deviation from the predictions of that
effective force is then an unambiguous signal of physics beyond the
Standard Model. We show that recent, state-of-the-art measurements of
parity-violating electron scattering (PVES) on nuclear targets
\cite{Ito:2003mr,Spayde:2003nr,Maas:2004ta,Maas:2004dh,Aniol:2005zf,Aniol:2005zg,Armstrong:2005hs,Acha:2006my}
yield a dramatic improvement in the accuracy with which we probe the
weak neutral-current sector of the Standard Model at low energy.

For our purposes, the relevant piece of the weak force which
characterises the virtual-exchange of a $Z^0$-boson between an electron
and an up or down quark can be parameterised by the constants,
$C_{1u(d)}$, which are defined through the effective four-point
interaction by \cite{Yao:2006px}
\begin{equation}
{\cal L}_{\rm NC}^{eq}=-\frac{G_F}{\sqrt{2}}\bar{e}\gamma_\mu\gamma_5e \sum_q C_{1q}\bar{q}\gamma^\mu q\,.
\label{eq:LSM}
\end{equation}
These effective couplings are known to high-precision within the
Standard Model, from precision measurements at the $Z$-pole \cite{unknown:2005em} and
evolution to the relevant low-energy scale
\cite{Marciano:1983ss,Erler:2003yk}.  There are also parity-violating
contributions arising from the lepton vector-current coupling to the
quark axial-vector-current, with couplings, $C_{2q}$, 
defined in a similar manner.
Although the PVES asymmetries are also dependent on the
$C_{2q}$'s, they cannot be extracted from these measurements without
input from nonperturbative QCD.

As summarized by the Particle Data Group (PDG)~\cite{Yao:2006px},
existing data, particularly the determination of atomic parity
violation in Cesium~\cite{Bennett:1999pd}, primarily constrains the
{\em sum} of the up and down quark ``charges'', $C_{1u} + C_{1d}$.
The analysis of the new high-precision PVES data presented here now
permits us to extract an independent experimental constraint on the
{\em difference}, $C_{1u} - C_{1d}$. Combining this constraint with
previous experimental results leads to a significant
improvement in the allowed range of values for $C_{1u}$ and $C_{1d}$.
This constraint is determined within the experimental uncertainties of
the electroweak structure of the proton.  The new range of values
allowed for these fundamental constants is consistent with the
predictions of the Standard Model and severely constrains relevant new
physics --- to a mass scale beyond $\sim$1--5~TeV.

Much of the current experimental interest in precision PVES
measurements on nuclear targets has been focussed on revealing the
strange-quark content of the nucleon.  Progress in revealing the
strangeness form factors has seen a dramatic improvement over the past
few years, with experimental results being reported by SAMPLE at
MIT-Bates~\cite{Ito:2003mr,Spayde:2003nr}, PVA4 at
Mainz~\cite{Maas:2004ta,Maas:2004dh} and the HAPPEX
\cite{Aniol:2005zf,Aniol:2005zg} and G0 \cite{Armstrong:2005hs}
Collaborations at Jefferson Lab. Depending on the target and kinematic
configuration, these measurements are sensitive to different linear
combinations of the strangness form factors, $G_E^s$ and $G_M^s$, and
the nucleon anapole form factor \cite{Haxton:1989ap,Zhu:2000gn}.
Recently, we reported a global analysis \cite{Young:2006jc} of these
measurements to extract all form factors from data.

Incorporating the new high-precision data, recently published by the
HAPPEX Collaboration \cite{Acha:2006my}, into our global analysis
\cite{Young:2006jc}, yields the most precise determination of the
strange-quark currents to date, namely (at $Q^2 =0.1\gev^2$)
$G_E^s=0.002\pm0.018$ and $G_M^s=-0.01\pm0.25$ (correlation
coefficient $-0.96$). Should one further impose constrain to theory
estimates for the anapole form factor \cite{Zhu:2000gn}, as discussed
below, these numbers shift by less than one standard deviation (with
$G_E^s=-0.011\pm0.016$ and $G_M^s=0.22\pm0.20$).  Nevertheless, with
the best fits constrained by data alone, we now ascertain that, at the
95\% confidence level (CL), strange quarks contribute less than 5\% of
the mean-square charge radius and less than 6\% of the magnetic moment
of the proton. This new result offers further support for the latest
theoretical quantum chromodynamics calculations
\cite{Leinweber:2004tc,Leinweber:2006ug}.

This determination of the strangeness form factors intimately relies
on the accurate knowledge of the low-energy electroweak parameters of
Eq.~\ref{eq:LSM}. Here we demonstrate that the latest PVES
measurements are sufficient to probe new physics by testing the
$Q^2$-evolution of the Standard Model.

Our global analysis of PVES measurements fits the world data with a
systematic expansion of the relevant form factors in powers of
$Q^2$. In this way one makes greatest use of the entire data set,
notably the extensive study of the dependence on momentum transfer
between $0.1$ and $0.3\gev^2$ by the G0 experiment
\cite{Armstrong:2005hs}. We now allow the two coupling constants,
$C_{1u}$ and $C_{1d}$, to be determined by the data.

Most of the PVES data has been measured on a hydrogen target. 
For small momentum transfer, in the forward-scattering limit, the
parity-violating asymmetry can be written as
\begin{equation}
A_{LR}^p \simeq A_0 \left[ Q_{\rm weak}^p Q^2 + B_4 Q^4 +\ldots \right]\,,
\end{equation}
where the overall normalisation is given by
$A_0=-G_\mu/(4\pi\alpha\sqrt{2})$. The leading term in this expansion
directly probes the weak charge of the proton, related to the quark
weak charges by $Q_{\rm weak}^p=G_E^{Zp}(0)=-2(2C_{1u}+C_{1d})$. (We
note that in our earlier analysis~\cite{Young:2006jc}, the full expressions
for the relevant asymmetries were written in terms of radiative
correction factors~\cite{Musolf:1993tb}, related by
$\xi_V^p=-2(2C_{1u}+C_{1d})$ and $\xi_V^n=-2(C_{1u}+2C_{1d})$). The
next-to-leading order term, $B_4$, is the first place that 
hadronic structure enters,
with the dominant source of uncertainty coming from the neutral-weak,
mean-square electric radius and magnetic moment. Under the assumption
of charge symmetry 
this uncertainty naturally translates to the knowledge of the
strangeness mean-square electric radius and magnetic moment. By
considering different phenomenological parameterisations of the
elastic form factors, we have confirmed that the potential uncertainties
from this source have a negligible impact on our final result.

\begin{figure}
\includegraphics[width=\columnwidth]{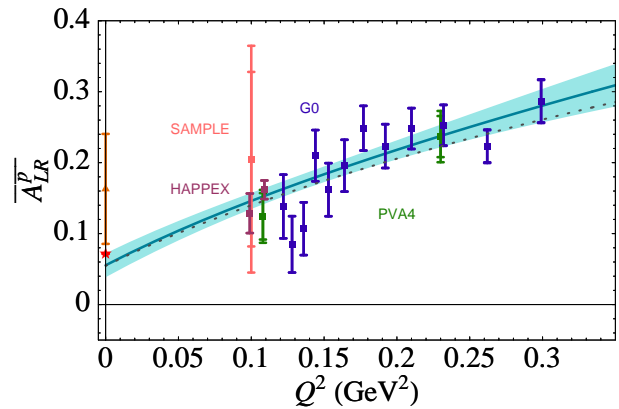}
\caption{Normalised, parity violating asymmetry 
measurements on a proton target, extrapolated to the forward-angle 
limit using our analysis of all world data on PVES (see text). The
extrapolation to $Q^2=0$ measures the proton's weak charge, where the
previous experimental knowledge (within uncertainties on the neutron
weak charge) is shown by the triangular data point, and the Standard
Model prediction by the star. The solid curve and shaded region
indicate, respectively, the best fit and 1-$\sigma$ bound, 
based upon our global fit to
all electroweak data. The dotted curve shows the resulting fit if one
incorporates the theoretical estimates~\cite{Zhu:2000gn} of the anapole
form factors of the nucleon.}
\label{fig:extrap}
\end{figure}

The extent of the data taken over the range $0.1<Q^2<0.3\gev^2$ allows
a reliable extrapolation in $Q^2$ to extract the proton's weak
charge. In Figure~\ref{fig:extrap} we show the various proton-target
measured asymmetries, extrapolated to zero degrees as explained below. 
The data is normalised as $\overline{A_{LR}^p} \equiv
A_{LR}^p/(A_0 Q^2)$, such that the intercept at $Q^2=0$ projects onto
$Q_{\rm weak}^p$.  The fit curve and uncertainty band is the result of
the full global fits, where helium, deuterium and all earlier relevant
neutral-weak current measurements \cite{Yao:2006px,Erler:2004cx} are
also incorporated.

Because each measurement has been performed at various scattering
angles, the data points displayed in Fig.~\ref{fig:extrap} have been
rotated to the forward-angle limit using the global fit of this
analysis, with the outer error bar on the data points indicating the
uncertainty arising from the $\theta\to 0$ extrapolation.  The
dominant source of uncertainty in this $\theta\to 0$ extrapolation
lies in the determination of the anapole form factor of the
nucleon. The experimentally-constrained uncertainty on the anapole
form factor is relatively large compared to estimates based upon
chiral perturbation theory supplemented with a vector meson dominance
model \cite{Zhu:2000gn}. Further constraining our fits to this
theoretical estimate yields the dotted curve in Fig.~\ref{fig:extrap},
where the discrepancy with the experimentally determined fit is always
less than one standard deviation --- and yields negligible impact on
the weak charge extraction. In particular,
the dotted curve changes our final lower-bound on the scale of new
physics, $\Lambda/g$, reported below, by less than 1 part
in $10^3$.

\begin{figure}
\includegraphics[width=\columnwidth]{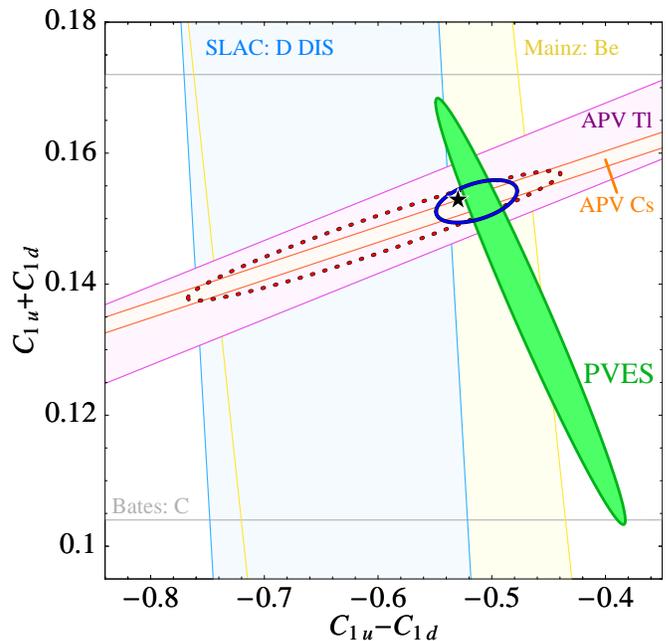}
\caption{Knowledge of the neutral weak effective couplings of
Eq.~(\ref{eq:LSM}).  The dotted contour displays the previous
experimental limits (95\% CL) reported in the PDG~\cite{Yao:2006px}
together with the prediction of the Standard Model (black star). The
filled ellipse denotes the new constraint provided by recent
high-precision PVES scattering measurements on hydrogen, deuterium and
helium targets (at 1 standard deviation), while the solid contour
(95\% CL) indicates the full constraint obtained by combining all
results.  All other experimental limits shown are displayed at 1
standard deviation.}
\label{fig:C1qNEW}
\end{figure}

This new constraint on the proton's weak charge provides an
essentially orthogonal constraint to combine with the precise atomic
parity-violation measurement on Cesium
\cite{Bennett:1999pd,Ginges:2003qt} --- which primarily constrains the
isoscalar combination of the weak quark charges.  The combined
analysis, which involves fitting both the hadronic structure
(strangeness and anapole) and electroweak parameters ($C_{1u,d}$,
$C_{2u,d}$), displays excellent agreement with the data, with a
reduced $\chi^2=21/23$.  The new improvement is displayed in
Fig.~\ref{fig:C1qNEW}, where the previous experimental knowledge is
summarised by the dotted ellipse \cite{Yao:2006px,Erler:2004cx}.  This
is to be compared with the solid contour obtained by combining all
earlier data with the powerful new constraint of the PVES data
(indicated by the filled ellipse). Although the principal constraint
from the PVES data is the proton's weak charge, the results are also
sensitive to the isoscalar combination of the $C_{1q}$ parameters
through the helium measurements --- resulting in the ellipse in
Fig.~\ref{fig:C1qNEW}.

The area of the allowed phase space of the effective couplings has
been reduced by a factor of 5, where the combined analysis produces
$C_{1u}+C_{1d}=0.1526\pm0.0013$ and $C_{1u}-C_{1d}=-0.513\pm0.015$
(correlation $+0.49$). Since these numbers are dominated by the atomic
parity violation and forward-angle PVES measurements, the correlations
with the $C_{2q}$ parameters are small.  Further, we note that the new
allowed region is in excellent agreement with the latest Standard
Model values \cite{Erler:2003yk,Yao:2006px},
$C_{1u}+C_{1d}=0.1529\pm0.0001$ and $C_{1u}-C_{1d}=-0.5297\pm0.0004$.

\begin{figure}
\includegraphics[width=\columnwidth]{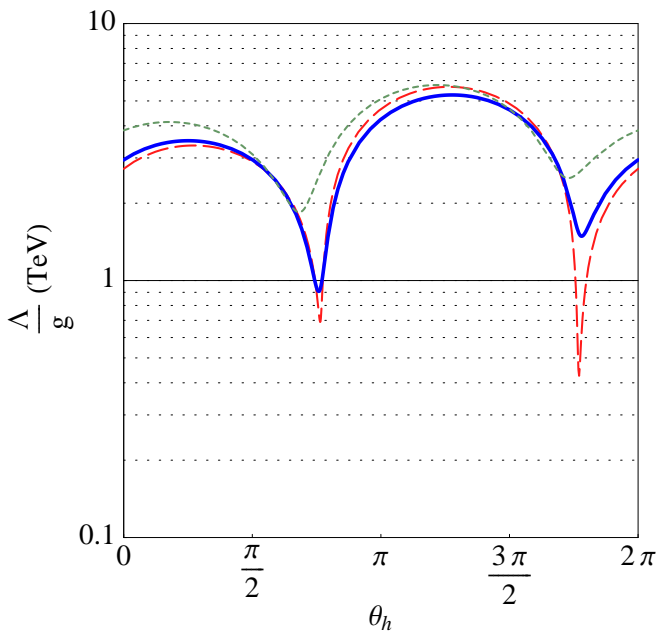}
\caption{Constraints on new physics beyond the Standard Model in terms
of the mass scale versus the flavour mixing angle. New physics is
ruled out at the 95\% CL below the curves.  The long-dashed curve shows
the limits before this work, corresponding to the dotted ellipse of
Fig.~\ref{fig:C1qNEW}. The solid curve shows the new limits based
on the solid contour in Fig~\ref{fig:C1qNEW}.  The short-dashed curve
displays the reach of the future proton weak charge measurement to be
performed at Jefferson Lab, {\em assuming} agreement with the Standard Model
is observed.}
\label{fig:ZPRIMES}
\end{figure}

The dramatic increase in precision shown by the solid contour in
{}Fig.~\ref{fig:C1qNEW} yields a new constraint on physics beyond the
Standard Model. Whatever the dynamical origin, new physics can be
expressed in terms of an effective contact interaction \cite{Erler:2003yk},
\begin{equation}
{\cal L}_{\rm NP}^{eq}=\frac{g^2}{\Lambda^2}\bar{e}\gamma_\mu\gamma_5 e \sum_q h_V^q \bar{q}\gamma^\mu q\,.
\end{equation}
With the characteristic energy scale, $\Lambda$, and coupling
strength, $g$, we set the isospin dependence by $h_V^u=\cos\theta_h$
and $h_V^d=\sin\theta_h$. In Fig.~\ref{fig:ZPRIMES} we display the
bounds on the size of such an interaction, where new physics is ruled
out at the 95\% CL below the curve. Whereas previous data constrained
a lower bound on relevant new physics $\Lambda/g>0.4\tev$, consistent
with direct searches at Fermilab, our new result lifts this limit to
$0.9\tev$.
Note that this limit applies to a generic class of quark-lepton
parity-violating contact interactions, arising from integrating out
all heavy, new-physics, degrees of freedom. The possibility of
cancellations between conspiring new physics at lower energies cannot
be ruled out by this analysis --- yet our results do then necessitate
such a cancellation for any proposed lower-energy new-physics
scenarios.  We further note that our determined limit involves
the ratio, $\Lambda/g$, hence these results directly imply a ``weak''
coupling for any potential discovery of new physics at a relatively
low energy.

To conclude, let us briefly summarize our main findings.  Our
systematic analysis of the world data on PVES has shown that the
effect of the hadronic form factors can be separated from the
low-energy, effective weak charges $C_{1u}$ and $C_{1d}$.  Combining
the resulting constraint with that obtained from the study of atomic
parity violation data, one finds an extremely tight range of allowed
values for {\bf both} $C_{1u}$ and $C_{1d}$ --- see the solid contour
in {}Fig.~\ref{fig:C1qNEW}. Not only is the result in excellent
agreement with the predictions of the Standard Model, but the
reduction in the range of allowed values of $C_{1u}$ and $C_{1d}$ is
such that it severely limits the possibilities of relevant new physics
(beyond the Standard Model) below a mass scale $\sim$1--5 TeV.

The success of this analysis makes evident the significance of what
can be achieved with future high-precision measurements, such as
Q-weak, the proton PVES experiment planned at Jefferson
Lab. Supplementing our current analysis with data of the accuracy
expected from that experiment (under the assumption of agreement with 
the Standard Model) yields a reduction of the allowed region
of $C_{1u,1d}$ by a further factor of 5. The potential impact of this
is shown in Fig.~\ref{fig:ZPRIMES}, where over most of the range of
$\theta_h$, this would raise the lower bound on the mass of a possible
$Z'$ by roughly $1\tev$. Of course, it is also possible that Q-weak 
could discover a deviation from the Standard Model which would 
constrain both the mass--coupling ratio and flavor dependence of the relevant 
new physics, such as a new $Z^\prime$. 
In the event of a discovery at the LHC, then measurements such
as Q-weak will play a key role in determining the characteristics of
the new interaction.

We wish to express our gratitude to K.~Carter and S.~Corneliussen for
helpful remarks on the manuscript.
This work was supported by DOE contract
DE-AC05-06OR23177, under which Jefferson Science Associates, LLC,
operates Jefferson Lab.

%%%%%%%%%%%%%%%%%%%%%%%%%%%%%%%%%%%%%%%%%%%%%%%%%%%%%%%%%%%%%%%%%%%

%

% Following is a new environment, {scilastnote}, that's defined in the
% preamble and that allows authors to add a reference at the end of the
% list that's not signaled in the text; such references are used in
% *Science* for acknowledgments of funding, help, etc.

%\begin{scilastnote}
%\item 
%This work was supported by DOE contract 
%DE-AC05-06OR23177, under which Jefferson Science Associates, LLC,  
%operates Jefferson Lab.
%\end{scilastnote}

\end{document}